\documentclass[a4paper,12pt,onecolumn,oneside,notitlepage,final]{article}
\usepackage[dvipdfmx]{graphicx}
\usepackage{amssymb} %
\usepackage{authblk}
\def\Journal#1#2#3#4{{#1} {\bf #2}, #3 (#4)}

\def\APJ{Astrophys. J.}

\def\IJMPA{Int. J. Mod. Phys. A}

\def\JHEP{JHEP}

\def\JETPUSSR{JETP (USSR)}
 
\def\MPLA{Mod. Phys. Lett. A}

\def\NPBSUPPL{Nucl. Phys. B. Proc. Suppl.}
\def\PLB{{Phys. Lett.} B}

\def\PRL{Phys. Rev. Lett.}
\def\PRD{Phys. Rev. D}

\def\PTP{Prog. Theor. Phys.}
\def\RMP{Rev. Mod. Phys.}

\def\SCIENCE{Science}

\def\ZETP{Zh. Eksp. Teor. Fiz.}

\def\Erratum{Erratum-ibid}

\begin{document}
\title{Parameterization of Pontecorvo-Maki-Nakagawa-Sakata mixing matrix based on CP-violating bipair neutrino mixing}

\author{\footnotesize Jun Iizuka, Teruyuki Kitabayashi\footnote{teruyuki@keyaki.cc.u-tokai.ac.jp}, Yuki Minagawa and Masaki Yasu\`{e}\footnote{yasue@keyaki.cc.u-tokai.ac.jp}}

\affil{Department of Physics, Tokai University,\\
4-1-1 Kitakaname, Hiratsuka, Kanagawa, 259-1292, Japan\\
}

\date{\small \today}
\maketitle

\begin{abstract}
CP violation in neutrino interactions is described by three phases contained in Pontecorvo-Maki-Nakagawa-Sakata mixing matrix ($U_{PMNS}$).  We argue that the phenomenologocally consistent result of the Dirac CP violation can be obtained if $U_{PMNS}$ is constructed along bipair neutrino mixing scheme, namely, requiring that $ \left| U_{12} \right| = \left| U_{32} \right|~{\rm and}~\left| U_{22} \right| = \left| U_{23} \right|~\left(\rm case~1\right)$ and $ \left| U_{12} \right| = \left| U_{22} \right|~{\rm and}~\left| U_{32} \right| = \left| U_{33} \right|~\left(\rm case~2\right)$, where $U_{ij}$ stands for the $i$-$j$ matrix element of $U_{PMNS}$. As a results, the solar, atmospheric and reactor neutrino mixing angles $\theta_{12}$, $\theta_{23}$ and $\theta_{13}$, respectively, are correlated to satisfy $\cos 2{\theta _{12}} = \sin^2\theta_{23} - \tan^2\theta_{13}$ (case 1) or $\cos 2{\theta _{12}} = \cos^2\theta_{23} - \tan^2\theta_{13}$ (case 2). Furthermore, if Dirac CP violation is observed to be maximal, $\theta_{23}$ is determined by $\theta_{13}$ to be: $\sin^2\theta_{23} \approx \left( {\sqrt 2  - 1} \right)\left( {\cos^2\theta_{13} + \sqrt 2 \sin^2\theta_{13}} \right)$ (case 1) or $\cos^2\theta_{23} \approx \left( {\sqrt 2  - 1} \right)\left( {\cos^2\theta_{13} + \sqrt 2 \sin^2\theta_{13}} \right)$ (case 2). For the case of non-maximal Dirac CP violation, we perform numerical computation to show relations between the CP-violating Dirac phase and the mixing angles.
\end{abstract}

\section{Introduction}
The neutrino oscillations have been experimentally confirmed by the Super-Kamiokande collaboration \cite{SK1,SK2,SK3,SK4}, who observed a deficit in the flux of atmospheric neutrinos. A similar oscillation phenomenon has been long suggested to occur in solar neutrinos \cite{OldSolor1,OldSolor2,OldSolor3,OldSolor4,OldSolor5,OldSolor6} and have been finally confirmed by various collaborations \cite{Sun1,Sun2,Sun3,Sun4,Sun5}.  Theoretically, the neutrino oscillations are realized if neutrinos have different masses and can be explained by mixings of three flavor neutrinos $\nu_{e,\mu,\tau}$: the $\nu_\mu$-$\nu_\tau$ mixing for the atmospheric neutrino oscillation and the $\nu_e$-$\nu_\mu$ mixing for the solar neutrino oscillation. These mixings are well described by a unitary matrix $U_{PMNS}$ \cite{PMNS1,PMNS2} involving three mixing angles $\theta_{12,23,13}$, which converts three massive neutrinos $\nu_{1,2,3}$ into $\nu_{e,\mu,\tau}$. Furthermore, leptonic CP violation is induced if $U_{PMNS}$ contains phases, which are given by one CP-violating Dirac phase $\delta$ and two CP-violating Majorana phases $\phi_{2,3}$ \cite{Majorana1,Majorana2,Majorana3}.  The standard parameterization of $U_{PMNS}$ \cite{PDG} is known to take the following form given by $U_{PMNS}=UK$ with
\begin{eqnarray}
U&=&\left( \begin{array}{ccc}
  c_{12}c_{13} &  s_{12}c_{13}&  s_{13}e^{-i\delta}\\
  -c_{23}s_{12}-s_{23}c_{12}s_{13}e^{i\delta}
                                 &  c_{23}c_{12}-s_{23}s_{12}s_{13}e^{i\delta}
                                 &  s_{23}c_{13}\\
  s_{23}s_{12}-c_{23}c_{12}s_{13}e^{i\delta}
                                 &  -s_{23}c_{12}-c_{23}s_{12}s_{13}e^{i\delta}
                                 & c_{23}c_{13}\\
\end{array} \right),
\nonumber \\
K &=& {\rm diag}(1, e^{i\phi_2/2}, e^{i\phi_3/2}),
\label{Eq:UuPDG}
\end{eqnarray}
where $c_{ij}=\cos\theta_{ij}$ and $s_{ij}=\sin\theta_{ij}$ and $\theta_{ij}$ represents a $\nu_i$-$\nu_j$ mixing angle ($i,j$=1,2,3). It is understood that the charged leptons and neutrinos are rotated, if necessary, to give diagonal charged-current interactions and to define the flavor neutrinos. 
The latest result of the experimental observation of three mixing angles is summarized as follows \cite{Nudata}:
\begin{eqnarray}
\sin^2 \theta_{12} &=& 0.304^{+0.013}_{-0.012}~({\rm NH~or~IH}),
\nonumber\\
\sin^2 \theta_{23} &=& 0.452^{+0.052}_{-0.028}~({\rm NH}), 0.579^{+0.025}_{-0.037}~({\rm IH}),
\nonumber\\
\sin^2 \theta_{13} &=& 0.0218^{+0.0010}_{-0.0010}~({\rm NH}), 0.0219^{+0.0011}_{-0.0010}~({\rm IH}),
\nonumber\\
\delta(^\circ)&=& 306^{+39}_{-70}~({\rm NH}),\quad 254^{+63}_{-62}~({\rm IH}),
\label{Eq:observedangle}
\end{eqnarray}
for the normal mass hierarchy (NH) or for the inverted mass hierarchy (IH).

There are various theoretical discussions that predict these mixing angles in literatures \cite{reviewObMixings1,reviewObMixings2,Kisslinger2013}.  Among others, original bipair neutrino mixing scheme has been proposed  \cite{pureBipair} and is based on the following constraints on $U_{PMNS}$: 
\begin{eqnarray}
&&
 \left| U_{12} \right| = \left| U_{32} \right|~{\rm and}~\left| U_{22} \right| = \left| U_{23} \right|~\left(\rm case~1\right),
\nonumber \\
&&
 \left| U_{12} \right| = \left| U_{22} \right|~{\rm and}~ \left| U_{32} \right| = \left| U_{33} \right|~\left(\rm case~2\right),
\label{Eq:org_bipair}
\end{eqnarray}
both with $U_{13}=0$, where $U_{ij}$ ($i,j$=1,2,3) stands for the $i$-$j$ matrix element of $U_{PMNS}$. It is found that
\begin{eqnarray}
\sin^2\theta_{23}&=&\sqrt{2}-1~(\approx 0.414)~\left(\rm case~1\right),
\nonumber \\
\sin^2\theta_{23}&=&2-\sqrt{2}~(\approx 0.586)~\left(\rm case~2\right),
\label{Eq:theta-org}
\end{eqnarray}
as well as $\sin^2\theta_{12}=1-1/\sqrt{2}(\approx 0.293)$ and $\sin^2\theta_{13}=0$.  It is clear that the case 1 can describe NH while the case 2 can describe IH. Since the observed value of $\theta_{13}$ turns out be clearly nonvanishing \cite{theta13_1,theta13_2,theta13_3,theta13_4,theta13_5,theta13_6,theta13_7,theta13_8,theta13_9,theta13_10}, there appears an interesting possibility to detect leptonic CP violation in neutrino interactions \cite{leptonicCP}.  To induce $\theta_{13}\neq 0$, we have discussed how contributions from charged leptons modify the predictions of the bipair neutrino mixing scheme, which slightly break the required conditions, and have estimated sizes of CP-violating Dirac and Majorana phases \cite{modifiedBipair1,modifiedBipair2,modifiedBipair3}. 

In this letter, we explore alternative possibility to estimate effects of leptonic CP violation. We retain the constraints on $U_{PMNS}$ intact even if sources of CP violation are included \cite{partial_mutau1,partial_mutau2}. Since effects from CP-violating Majorana phases are hidden, we expect a certain correlation of the mixing angles to CP-violating Dirac phase to be clarified.

\section{CP-violating bipair neutrino mixing}
CP-violating bipair neutrino mixing necessarily contains complex-valued $U_{ij}$. It is, therefore, reasonable to require the following bipair constraints in Eq.(\ref{Eq:org_bipair}) extended to include complex $U_{ij}$:
\begin{eqnarray}
 \left| U_{12} \right| &=& \left| U_{32} \right|~{\rm and}~\left| U_{22} \right| = \left| U_{23} \right|~\left(\rm case~1\right),
\nonumber \\
 \left| U_{12} \right| &=& \left| U_{22} \right|~{\rm and}~\left| U_{32} \right| = \left| U_{33} \right|~\left(\rm case~2\right).
\label{Eq:exact_bipair}
\end{eqnarray}
When the nonvanishing $\theta_{13}$ and $\delta$ are taken into account in the requirement to obtain the CP-violating bipair neutrino mixing, Eq.(\ref{Eq:UuPDG}) gives
\begin{eqnarray}
c_{13}^2 - c_{23}^2s_{13}^2 - s_{23}^2 &= &\left( {s_{23}^2 + c_{13}^2 - c_{23}^2s_{13}^2} \right)\cos 2{\theta _{12}}  + {s_{13}}\sin 2{\theta _{12}}\sin 2{\theta _{23}}\cos {\delta},  
\label{Eq:first}
\end{eqnarray}
from $\left| {{U_{12}}} \right| = \left| {{U_{32}}} \right|$, and 
\begin{eqnarray}
2s_{23}^2c_{13}^2 - c_{23}^2 - s_{23}^2s_{13}^2 &=&\left( {c_{23}^2 - s_{23}^2s_{13}^2} \right)\cos 2{\theta _{12}} - {s_{13}}\sin 2{\theta _{12}}\sin 2{\theta _{23}}\cos {\delta},
\label{Eq:second}
\end{eqnarray}
from $\left| {{U_{22}}} \right| = \left| {{U_{23}}} \right|$, for the case 1.  Results corresponding to the case 2 are obtained by the interchange of $c_{23}\leftrightarrow s_{23}$ simultaneously with the replacement of $s_{13}\rightarrow -s_{13}$.  
It is readily observed that the simple sum of Eqs.(\ref{Eq:first}) and (\ref{Eq:second}) yields 
\begin{eqnarray}
\cos 2{\theta _{12}} &=& \sin^2\theta_{23} - \tan^2\theta_{13} ~\left(\rm case~1\right), \nonumber \\
\cos 2{\theta _{12}} &=& \cos^2\theta_{23} - \tan^2\theta_{13}~\left(\rm case~2\right),
\label{Eq:exactbipair_predictions}
\end{eqnarray}
which represents a unique prediction of the CP-violating bipair neutrino mixing. The experimental data in Eq.(\ref{Eq:observedangle}) well satisfy the relations in Eq.(\ref{Eq:exactbipair_predictions}). For instance, $\sin^2 \theta_{12}$ is predicted to be $\sin\theta_{12} = 0.300$ for given values of $\sin^2 \theta_{23} = 0.425$ and $\sin^2 \theta_{13} = 0.0244$ (case 1) or of $\sin^2 \theta_{23} = 0.576$ and $\sin^2 \theta_{13} = 0.0234$ (case 2).

For the practical purpose, we may safely omit terms proportional to $s^3_{13}$ because of the smallness of $s^2_{13}$ as in Eq.(\ref{Eq:observedangle}). The simplest calculation can be done if $\cos\delta=0$ indicating maximal CP violation is taken; thereby, $\delta = 3\pi/2$ to be consistent with Eq.(\ref{Eq:observedangle}) and we obtain that   
\begin{eqnarray}
s_{23}^2 &\approx& \left( {\sqrt 2  - 1} \right)\left( {c_{13}^2 + \sqrt 2 s_{13}^2} \right)(\equiv {\hat s}_{23}^2)~\left(\rm case~1\right), \nonumber \\
c_{23}^2 &\approx& \left( {\sqrt 2  - 1} \right)\left( {c_{13}^2 + \sqrt 2 s_{13}^2} \right)(\equiv {\hat c}_{23}^2)~\left(\rm case~2\right),
\label{Eq:s23-maximal}
\end{eqnarray}
where the value of $\theta_{23}$ evaluated at $\cos\delta=0$ is to be denoted by ${\hat \theta}_{23}$ giving ${\hat s}_{23}=\sin {\hat \theta}_{23}$ and so on.  When $\cos\delta\neq 0$, Eqs.(\ref{Eq:first}) and (\ref{Eq:second}) indicate that corrections to ${\hat \theta}_{23}$ accompany the factor ${s_{13}}\cos {\delta}\sin 2{\theta _{12}}$,  As a result, we find that 
\begin{eqnarray}
s_{23}^2 = \hat s_{23}^2 + \Delta,
\label{Eq:s23-arbitrary}
\end{eqnarray}
where
\begin{eqnarray}
\Delta  &=& -\frac{{{s_{13}}\cos {\delta}\sin 2{\theta _{12}}}}{2} \left[ \frac{{\sin 2{{\hat \theta }_{23}}}}{{1 + \hat s_{23}^2}} - {s_{13}}\cos {\delta}\sin 2{\theta _{12}}\left( {\frac{{\cos 2{{\hat \theta }_{23}}}}{{{{\left( {1 + \hat s_{23}^2} \right)}^2}}} - \frac{{{{\sin }^2}2{{\hat \theta }_{23}}}}{{4{{\left( {1 + \hat s_{23}^2} \right)}^3}}}} \right) \right] \nonumber, \\
\label{Eq:s23-Delta}
\end{eqnarray}
for the case 1.  The final expression is obtained by neglecting ${\mathcal O}(s^3_{13})$ after Eq.(\ref{Eq:s23-Delta}) is expanded in series of $s^2_{13}$. The interchange of ${\hat c}_{23}\leftrightarrow {\hat s}_{23}$ with the replacement of $s_{13}\rightarrow -s_{13}$ gives the result for the case 2. 

\begin{figure}[t]
\begin{center}
\includegraphics{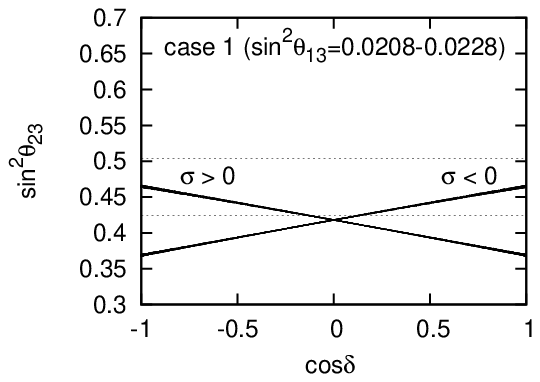}
\includegraphics{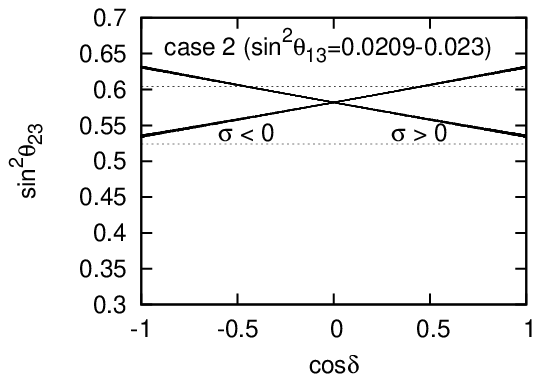}
\caption{Prediction of $\sin^2\theta_{23}$ as a function of $\cos\delta$ for the case 1 (left) or for the case 2  (right). Each region sandwiched by two dotted horizontal lines indicates the experimentally allowed region of $\sin^2\theta_{23}$.}
\label{Fig:theta23_delta}
\end{center}
\end{figure}

\begin{figure}[t]
\begin{center}
\includegraphics{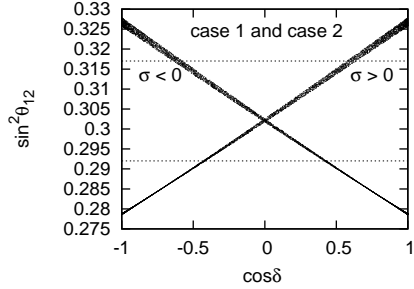}
\caption{The same as in FIG.\ref{Fig:theta23_delta} but for $\sin^2\theta_{12}$.}
\label{Fig:theta12_delta}
\end{center}
\end{figure}

\section{CP-violating Dirac phase}
To visually see the dependence of the mixing angles on $\delta$, we perform numerical analysis. Since theoretical predictions depend on $\delta$ in the form of $\cos\delta$, we use $\cos\delta$ instead of $\delta$ to draw figures. 

The figures Fig.\ref{Fig:theta23_delta} and \ref{Fig:theta12_delta} describe how $\sin^2\theta_{12,23}$ vary with $\delta$ as a function of $\cos\delta$ using the observed data of $\sin^2\theta_{13}$, where two different lines depend on the sign of $s_{13}\sin 2\theta_{12}\sin 2\theta_{23}$ in $\Delta$ denoted by $\sigma$.  When $\sin^2\theta_{12,23}$ are constrained to satisfy the observed data of Eq.(\ref{Eq:observedangle}), the range of $\delta$ can be determined to be:

\begin{itemize}
\item for the case 1 (left in Fig.\ref{Fig:theta23_delta}), $-1\lesssim\cos\delta\lesssim -0.11$ ($\sigma > 0$) or $0.11\lesssim\cos\delta\lesssim 1$ ($\sigma < 0$),
\item for the case 2 (right in Fig.\ref{Fig:theta23_delta}), $-0.46\lesssim\cos\delta\lesssim 1$ ($\sigma > 0$) or $-1\lesssim\cos\delta\lesssim 0.46$ ($\sigma < 0$),
\item for the both cases (Fig.\ref{Fig:theta12_delta}), $-0.42\lesssim\cos\delta\lesssim 0.61$ ($\sigma > 0$) or  $-0.61\lesssim\cos\delta\lesssim 0.42$ ($\sigma < 0$).
\end{itemize}
There are other ranges covered by $2\pi-\delta$ giving the same value of $\cos\delta$. By combining the above results, we find the predicted ranges of $\cos\delta$:
\begin{itemize}
\item for the case 1, $-0.42\lesssim\cos\delta\lesssim -0.11$ ($\sigma > 0$) or $-0.11\lesssim\cos\delta\lesssim 0.42$ ($\sigma < 0$),
\item for the case 2, $-0.42\lesssim\cos\delta\lesssim 0.61$ ($\sigma > 0$) or $-0.61\lesssim\cos\delta\lesssim 0.42$ ($\sigma < 0$),
\end{itemize}
It is obvious that our predictions are consistent with the observed data of $\delta$.

\section{Summary}
We have advocated that the CP-violating bipair neutrino mixing scheme well describes the observed property of neutrinos. The simplest relation among the mixing angles is found to be: $\cos 2{\theta _{12}} = \sin^2\theta_{23} - \tan^2\theta_{13}$ for the case 1 or $\cos 2{\theta _{12}} = \cos^2\theta_{23} - \tan^2\theta_{13}$ for the case 2. Furthermore, if Dirac CP violation is observed to be maximal, $\sin^2\theta_{23}$ is determined by $\sin^2\theta_{13}$ to be: $\sin^2\theta_{23} \approx \left( {\sqrt 2  - 1} \right)\left( {\cos^2\theta_{13} + \sqrt 2 \sin^2\theta_{13}} \right)$ for the case 1 or $\cos^2\theta_{23} \approx \left( {\sqrt 2  - 1} \right)\left( {\cos^2\theta_{13} + \sqrt 2 \sin^2\theta_{13}} \right)$ for the case 2. 

It is emphasised that the CP-violating bipair neutrino mixing predicts the experimentally favored $\sin^2\theta_{23}>0.5$ for the inverted mass hierarchy to be around $\sin^2\theta_{23}=2-\sqrt{2}$. Although the CP-violating bipair neutrino mixing does not originate from any symmetry argument imposed either on the neutrino mass matrix or on the Lagrangian, the predicted values of the neutrino mixing angles and the CP-violating Dirac phase are well compatible with the observed data.  What is the origin of the CP-violating bipair neutrino mixing scheme will remain an issue for future investigations.



\begin{thebibliography}{99}
\bibitem{SK1}
Y. Fukuda {\it et al.}, [Super-Kamiokande Collaboration], \Journal{\PRL}{81}{1562}{1998}.

\bibitem{SK2}
Y. Fukuda {\it et al.}, [Super-Kamiokande Collaboration], \Journal{\PRL}{82}{2430}{1999}.

\bibitem{SK3}
T. Kajita, \Journal{\NPBSUPPL}{77}{123}{1999}.

\bibitem{SK4}
T. Kajita and Y. Totsuka, \Journal{\RMP}{73}{85}{2001}.

\bibitem{OldSolor1}
    J.N. Bahcall, W.A. Fowler, I. Iben and R.L. Sears, \Journal{\APJ}{137}{344}{1963}.

\bibitem{OldSolor2}
    J. Bahcall, \Journal{\PRL}{12}{300}{1964}.

\bibitem{OldSolor3}
    R. Davis, Jr., \Journal{\PRL}{12}{303}{1964}.

\bibitem{OldSolor4}
    R. Davis, Jr., D.S. Harmer and K.C. Hoffman, \Journal{\PRL}{20}{1205}{1968}.

\bibitem{OldSolor5}
    J.N. Bahcall, N.A. Bahcall and G. Shaviv, \Journal{\PRL}{20}{1209}{1968}.

\bibitem{OldSolor6}
    J.N. Bahcall and R. Davis, Jr., \Journal{\SCIENCE}{191}{264}{1976}.

\bibitem{Sun1}
	Y. Fukuda {\it et al.}, [Super-Kamiokande Collaboration], \Journal{\PRL}{81}{1158}{1998}; [\Journal{\Erratum}{81}{4279}{1998}].

\bibitem{Sun2}
	B.T. Cleveland {\it et al.}, \Journal{\APJ}{496}{505}{1998}.

\bibitem{Sun3}
	W. Hampel {\it et al.}, [GALLEX Collaboration],  \Journal{\PLB}{447}{127}{1999}.

\bibitem{Sun4}
	Q.A. Ahmad {\it et al.}, [SNO Collaboration], \Journal{\PRL}{87}{071301}{2001}.

\bibitem{Sun5}
	Q.A. Ahmad {\it et al.}, [SNO Collaboration], \Journal{\PRL}{89}{011301}{2002}.

\bibitem{PMNS1} 
	B. Pontecorvo, \Journal{\JETPUSSR}{7}{172}{1958} [\Journal{\ZETP}{34}{247}{1958}].

\bibitem{PMNS2} 
	Z. Maki, M. Nakagawa and S. Sakata, \Journal{\PTP}{28}{870}{1962}. 

\bibitem{Majorana1}
	S.M. Bilenky, J. Hosek and S.T. Petcov, \Journal{\PLB}{94}{495}{1980}.

\bibitem{Majorana2}
	J. Schechter and J.W.F. Valle, \Journal{\PRD}{22}{2227}{1980}.

\bibitem{Majorana3}
	M. Doi, T. Kotani, H. Nishiura, K. Okuda and E. Takasugi, \Journal{\PLB}{102}{323}{1981}.

\bibitem{PDG} 
J. Beringer et al. (Particle Data Group), \Journal{\PRD}{86}{010001}{2012}.

\bibitem{Nudata}
M.C. Gonzalez-Garcia, M. Maltoni, and T. Schwetz, \Journal{\JHEP}{1411}{052}{2014}.

\bibitem{reviewObMixings1}
G. Altarelli and F. Feruglio, \Journal{\RMP}{82}{2701}{2010}.

\bibitem{reviewObMixings2}
G. Altarelli, \Journal{\IJMPA}{29}{1444002}{2014}.

\bibitem{Kisslinger2013}
L. S. Kisslinger, \Journal{\MPLA}{28}{1350153}{2013}.

\bibitem{pureBipair}
T. Kitabayashi and M. Yasu\`{e}, \Journal{\PLB}{696}{478}{2011}.

\bibitem{theta13_1}
K. Abe {\it et al.}, [T2K Collaboration], \Journal{\PRL}{107}{041801}{2011}.

\bibitem{theta13_2}
K. Abe {\it et al.}, [T2K Collaboration], \Journal{\PRL}{111}{211803}{2013}.

\bibitem{theta13_3}
K. Abe {\it et al.}, [T2K Collaboration], \Journal{\PRL}{112}{061802}{2014}.

\bibitem{theta13_4}
P. Adamson {\it et al.}, [MINOS Collaboration], \Journal{\PRL}{107}{181802}{2011}.

\bibitem{theta13_5}
P. Adamson {\it et al.}, [MINOS Collaboration], \Journal{\PRL}{110}{171801}{2013}.

\bibitem{theta13_6}
P. Adamson {\it et al.}, [MINOS Collaboration],  \Journal{\PRL}{110}{251801}{2013}.

\bibitem{theta13_7}
Y. Abe {\it et al.},  [Double Chooz Collaboration], \Journal{\PRD}{86}{052008}{2012}.

\bibitem{theta13_8}
J. K. Ahn {\it et al.}, [RENO Collaboration], ''Observation of Reactor Electron Antineutrino Disappearance in the RENO Experiment", arXiv:1204.0626 [hep-ex].

\bibitem{theta13_9}
F. P. An {\it et al.}, [DAYA-BAY Collaboration], \Journal{\PRL}{108}{171803}{2012}.

\bibitem{theta13_10}
F. P. An {\it et al.}, [DAYA-BAY Collaboration], \Journal{\PRL}{112}{061801}{2014}.

\bibitem{leptonicCP}
See for example, H. Minakata, ``Neutrino Physics Now and in the Near Future", arXiv:1403.3276 [hep-ph].

\bibitem{modifiedBipair1}
T. Kitabayashi and M. Yasu\`{e}, \Journal{\PLB}{713}{206}{2012}.

\bibitem{modifiedBipair2}
T. Kitabayashi and M. Yasu\`{e}, \Journal{\PLB}{726}{356}{2013}.

\bibitem{modifiedBipair3}
J. Iizuka, Y. Kaneko, T. Kitabayashi, N. Koizumi and M. Yasu\`{e}, \Journal{\PLB}{732}{191}{2014}.

\bibitem{partial_mutau1}
H. Qu and Bo-Q. Ma, \Journal{\PRD}{88}{037301}{2013}.

\bibitem{partial_mutau2}
Z.-z. Xing and S. Zhou, \Journal{\PLB}{737}{196}{2014}.

\end{thebibliography}
\end{document}